\documentclass{PoS}

\def\bea{\begin{eqnarray}}
\def\eea{\end{eqnarray}}
\def\nn{\nonumber}
\def\Be{{}^8\textrm{Be}}

\title{Explanation of the Beryllium Anomaly in a $U(1)'$-Extended 2-Higgs Doublet Model}

\ShortTitle{The Beryllium anomaly in a $U(1)'$-2HDM}

\author{\speaker{Luigi Delle Rose}\\
        School of Physics and Astronomy, University of Southampton, Highfield, Southampton SO17 1BJ, UK\\
        Particle Physics Department, Rutherford Appleton Laboratory, Chilton, Didcot, Oxon OX11 0QX, UK \\
        E-mail: \email{l.delle-rose@soton.ac.uk}}

\author{Shaaban Khalil\\
        Center for Fundamental Physics, Zewail City of Science and Technology, 6 October City, Giza 12588, Egypt\\
        E-mail: \email{skhalil@zewailcity.edu.eg}}

\author{Stefano Moretti\\
        School of Physics and Astronomy, University of Southampton, Highfield, Southampton SO17 1BJ, UK\\
        Particle Physics Department, Rutherford Appleton Laboratory, Chilton, Didcot, Oxon OX11 0QX, UK\\
        E-mail: \email{s.moretti@soton.ac.uk}}

\abstract{
We consider an extension of the Standard Model with an extra gauged $U(1)'$ symmetry in presence of a 2-Higgs Doublet Model structure of the scalar sector. 
We show that this scenario is able to explain the anomaly observed by the Atomki collaboration in the decay of an excited state of Beryllium via a light spin-1 $Z'$ with a mass of 17 MeV.
}

\FullConference{EPS-HEP 2017, European Physical Society conference on High Energy Physics\\
		5-12 July 2017\\
		Venice, Italy}

\begin{document}

\section{The Beryllium anomaly in a $U(1)'$-2HDM}

\begin{figure}
\centering
\includegraphics[width=7.cm,height=7.cm]{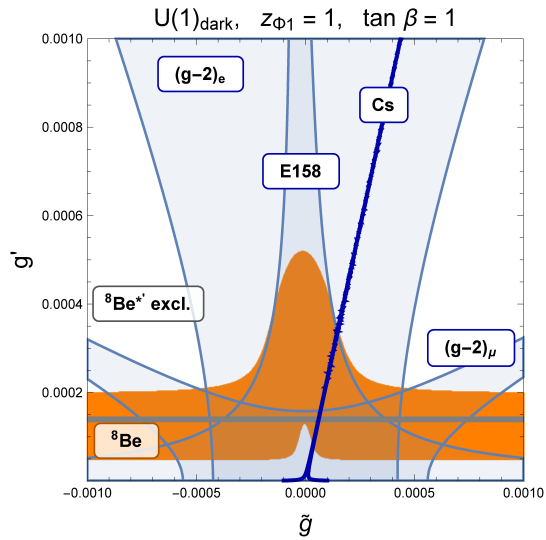}
\caption{
The orange region is the allowed parameter space explaining the Beryllium anomaly while the white region above is excluded by the non-observation of the $Z'$ in the ${\Be^{*}}'$ transition. 
The other shaded regions represent the parameter space allowed by the M{\o}ller scattering and the $g-2$ of electron and muon.
The blue line represents values of $g'$ and $\tilde g$ complying with atomic parity violation in Cs. 
The horizontal grey stripe determines the values of $g'$ for which the $Z'$ mass is generated by the electroweak vev.
\label{fig:typeI}}
\end{figure}

The Atomki pair spectrometer experiment \cite{Krasznahorkay:2015iga} has 
studied the decay of the excited $^8{\rm Be}$ nuclei, focusing on the $e^+e^-$ internal pair creation process. 
These excitations were produced with a beam of protons directed on a target of Lithium ($^7{\rm Li}$). The possibility to tune the energy of the protons with high accuracy 
has allowed for a resonant production and precise selection of the different $^8{\rm Be}$ excitations. 
An anomaly has been observed in the decay of the energy level characterised by an excitation energy of 18.15~MeV, the $^8{{\rm Be}^*}$ state (with
spin-parity $J^P=1^+$ and isospin $T=0$), into the ground state $^8{\rm Be}$ ($J^P=0^+$ and $T=0$).
From the analysis of the properties of the electron-positron pair, in particular of their opening angle and invariant mass distributions, the Atomki collaboration has observed 
an excess consistent with the on-shell emission of an intermediate boson $X$ eventually decaying into $e^+e^-$.
The best fit to the mass $M_X$ and the corresponding normalised Branching Ratio (BR) are given by 
\bea
&& M_X = 16.7 \pm 0.35\ \textrm{(stat)}\ \pm 0.5\ \textrm{(sys)\ MeV}, \nn \\
&& \frac{{\rm BR}(^8{{\rm Be}^*} \to X + {^8{\rm Be}})}{{\rm BR}(^8{{\rm Be}^*} \to \gamma + {^8{\rm Be}})} \times {\rm BR}(X\to e^+ e^-) 
= 5.8 \times 10^{-6} \,,
\eea
with a statistical significance of the excess of about $6.8\,\sigma$ \cite{Krasznahorkay:2015iga}. \\
The same collaboration has also recently presented evidences of an excess, compatible with a 17 MeV boson mediation, in the ${\Be^{*}}'$ ($J^P=1^+$ and $T=1$) transition 
which is characterised by an excitation energy of 17.64~MeV \cite{Krasznahorkay:2015iga}. Because of the less available phase space, the anomaly in this decay channel is expected to be kinemetically suppressed.
The result is not public yet we do not account for it in the following analysis, but we ought to mention it. \\
An attempt to explain the properties of the $X$ boson was carried out in \cite{Feng:2016jff,Feng:2016ysn} in which a new spin-1 boson, $Z'$, with vector-like couplings
to Standard Model (SM) leptons and quarks, was considered. The couplings of such a light state with quarks are, in general, strongly constrained from  $\pi^0\to Z'+\gamma$ 
searches at the NA48/2 experiment \cite{Batley:2015lha}. Complying with this bound requires the $Z'$ interactions to quarks to satisfy the `protophobic' condition,  
namely, $|2{\epsilon}_u+{\epsilon}_d| \lesssim 10^{-3}$ where ${\epsilon}_u$ and ${\epsilon}_d$ are the couplings to up and down quarks normalised with the
positron charge $e$.
In the footsteps of these works, further studies of such models
have been performed in \cite{Gu:2016ege,Chen:2016dhm,Liang:2016ffe,Jia:2016uxs,Kitahara:2016zyb,Chen:2016tdz,Seto:2016pks,
Neves:2016ugb,Chiang:2016cyf}. 
A completely alternative explanation was proposed in \cite{Ellwanger:2016wfe} in which the $X$ boson was identified with a light pseudoscalar and its couplings to up and down type quarks
were qauntified to be about 0.3 times those of the SM Higgs. \\
In \cite{DelleRose:2017xil} we considered the impact of the Beryllium anomaly on a rather generic extension of the SM described by an extra $U(1)'$ group in which the $Z'$ interactions are characterised by a general $V/A$ structure, namely $J^\mu_{Z'} = \sum_f \bar \psi_f \gamma^\mu \left( C_{f, V} \gamma^\mu + C_{f, A} \gamma^\mu \gamma^5 \right) \psi_f$.
Such scenarios with a light gauge mediator have been extensively studied in the past literature \cite{Fayet:1980rr,Fayet:1980ad,Fayet:1990wx,Fayet:2007ua,Fayet:2008cn,Fayet:2016nyc}.  \\
Due to the presence of two Abelian gauge groups, $U(1)_Y \times U(1)'$, a kinetic mixing between the corresponding gauge fields is allowed and a new free parameter, $\tilde g$, is introduced beside 
the usual gauge coupling $g'$. In the limit in which  $g', \tilde g \ll 1$ the $Z'$ interactions are described by the coefficients
\bea
\label{eq:CVA_expanded}
C_{f, V} &\simeq&    \tilde g  c_W^2 \, Q_f + g'  \left[ z_\Phi (T^3_f - 2 s_W^2 Q_f)  + z_{f,V} \right] \,, \nn \\
C_{f, A} &\simeq&  g' \left[   -  z_\Phi \, T^3_f  +   z_{f,A} \right],
\eea
where $z_{f,V/A}$ are the vector and axial-vector $U(1)'$ charges, $s_w$ and $c_w$ are, respectively, the sin and cos of the Weinberg angle and $Q_f$, $T^3_f$ are the electromagnetic charge and 
the third component of the weak ispospin. $z_\Phi$ is the $U(1)'$ of the Higgs scalar or a combination of charges in a general scenario with more than one $SU(2)$ doublet. 
As shown in \cite{Kahn:2016vjr,DelleRose:2017xil}, the axial coefficients $C_{f,A}$ are suppressed with respect to the vector-like ones as a consequence of the gauge invariance of the 
Yukawa interactions and of the presence of a single $SU(2)$ Higgs doublet in the scalar sector (which impose relations between the $U(1)'$ charges of the SM fermions and the Higgs). 
Interestingly, this feature is not affected by the requirement of anomaly cancellation and by the presence of extra matter which could be necessary to account for it. 
On the other hand, the vector and axial-vector couplings of the $Z'$ are of the same order of magnitude in a scenario with multiple $SU(2)$ scalar doublets such as the two Higgs doublet model (2HDM) which has been considered here.
Indeed, $z_\Phi = z_{\Phi_1} \cos^2 \beta +  z_{\Phi_2} \sin^2 \beta$, where $\tan \beta$ is defined, as usual, as the ratio of the vevs, and the cancellation between the two terms in $C_{f, A}$ of Eq.(\ref{eq:CVA_expanded}) is not realised regardless of the gauge invariance of the Yukawa Lagrangian. \\
Before moving to the discussion of the anomaly in the Beryllium decay, we briefly comment on the spontaneous symmetry breaking in this $U(1)'$-2HDM configuration.
In particular, in the $g', \tilde g \ll 1$ limit, the mass of the $Z'$ is given by
\bea
\label{eq:Zpmass}
M_{Z'}^2 \simeq   m_{B'}^2 +  \frac{v^2}{4} {g'}^2 (z_{\Phi_1} - z_{\Phi_2})^2 \sin^2 (2 \beta),
\eea
where we have allowed for a possible mass source $m_{B'}$ from an extra SM-singlet scalar. Notice that, $M_{Z'}$ is non-zero even when $m_{B'} \rightarrow 0$ due to a difference between the $U(1)'$ charges $z_{\Phi_1}$ and $z_{\Phi_2}$ of the two scalar doublets. When $m_{B'} = 0$, one obtains for $M_{Z'} \simeq 17$ MeV and $v \simeq 246$ GeV, $g' \sim 10^{-4}$. 
Notice that, in the case of only one Higgs doublet, $M_{Z'} \simeq m_{B'}$ and the extra scalar degree of freedom is mandatory for the $Z'$ to acquire a mass. \\
The scenario described above is able to explain the excess in the $\Be^*$ decay through a $Z'$ resonance. 
In particular, the contribution of the axial-vector couplings in the $\Be^* \rightarrow \Be \, Z'$ transition is found to be proportional to $k/M_{Z'} \ll 1$ (with $k$ being the $Z'$ momentum) while the vector contribution is suppressed by $k^3/M_{Z'}^3$ \cite{Feng:2016ysn}. Indeed, in the first case the $Z'$ is emitted in a s-wave configuration while in the latter a p-wave configuration is realised.
Therefore, being $C_{f,V}$ and $C_{f,A}$ of the same order of magnitude, we can safely neglect the vector-like contribution. \\
For the sake of simplicity, we consider type-I 2HDM scenario accompanied by a $U(1)_\textrm{dark}$ configuration with $z_{f} =0$ and $z_{\Phi_2} = 0$ and we choose $z_{\Phi_2} = 1$ and $\tan \beta = 1$. 
Similar results may be obtained for different values of $\tan \beta$ and $z$. The parameter space compatible with the observed Atomki anomaly, 
together with the most constraining experimental bounds, is shown in Fig.~\ref{fig:typeI}. The results have been obtained using the matrix elements for the $\Be^*$ transition computed in \cite{Kozaczuk:2016nma}.
The orange region delineates the portion of parameter space complying with the best-fit of the $\Be^*$ decay rate in the mass range $M_{Z'} = 16.7 \, {\rm MeV} - 17.6 \, {\rm MeV}$ \cite{Krasznahorkay:2015iga,Feng:2016ysn} taking into account the uncertainties of the nuclear matrix elements \cite{Kozaczuk:2016nma}. 
The recent hints of an analogous transition in the isovector excitation ${\Be^{*}}'$ \cite{Krasznahorkay:2015iga} has not been considered here and the white region above the orange one accounts for the non-observation of the latter.
Finally, the horizontal grey stripe locates the values of $g'$ for which the mass of $Z'$ is entirely generated by the electroweak symmetry breaking triggered by the two Higgs doublets ($m_{B'} \simeq 0$).
The most relevant experimental bounds (shaded regions are allowed) are shown on the same plot. 
In particular, the strongest constraint comes from atomic parity violation in Cesium (Cs), 
namely from the measurement of the corresponding weak nuclear charge $\Delta Q_W$ \cite{Davoudiasl:2012ag,Bouchiat:2004sp},
required to satisfy $| \Delta Q_W | \lesssim 0.71$ at $2\sigma$ \cite{Porsev:2009pr}.
This bound can be safely satisfied if the $Z'$ has either only vector or axial-vector couplings. On the other hand, in a general scenario it imposes severe constraints on the ratio of the two gauge couplings $g',\tilde g$.
Moreover, the parity-violating M{\o}ller scattering measured at SLAC E158 \cite{Anthony:2005pm} has been enforced, which requires $|C_{e,V} C_{e,A}| \lesssim 10^{-8}$ for $M_{Z'} \simeq 17$ MeV \cite{Kahn:2016vjr},
together with the constraints from the anomalous magnetic moment of the electron and the muon \cite{Giudice:2012ms,Altmannshofer:2016brv,Bennett:2006fi,Blum:2013xva,Lindner:2016bgg}.
Neutrino-electron scattering processes may also impose severe constraints \cite{Vilain:1994qy,Deniz:2009mu,Bellini:2011rx} on the combination of $C_{e, V/A}$ and $C_{\nu,V}$ couplings such as the one obtained by the 
TEXONO collaboration \cite{Deniz:2009mu}. Indeed, in the protophobic scenario with a $Z'$ with only vector interactions, the constrained neutrino coupling of the $Z'$ is in tension with the anomalous $\Be^*$ decay rate,
since $C_{\nu,V} = -2 C_{n,V}$, where $C_{n,V}$ is the $Z'$ coupling to neutrons responsible for the explanation of the Atomki anomaly. 
Nevertheless, this situation is alleviated in the general case where the $Z'$ boson has also axial-vector interactions because the gauge couplings $g',\tilde g$ required to explain the anomaly are typically smaller than the ones of the protophobic scenario.
\section{Conclusions}
The model that we have considered above has the necessary features to explain with a light $Z'$ the anomalous decay of the excited state of the Beryllium. The presence of two Higgs doublets allows to 
exploit the contributions of both the vector and axial-vector couplings of the $Z'$ interactions which are useful, for instance, to alleviate the tension from the bounds on the electron-neutrino scattering.
Moreover, the values of the gauge coupling $g'$ which reproduce the excess in the $\Be^*$ decay also ensure that the mass of the $Z'$ gauge boson may be generated 
from the symmetry breaking of the electroweak sector of the SM, namely from the EW scale $v \simeq 246$ GeV.

\section*{Acknowledgements}
The work of LDR and SM is supported in part by the NExT Institute. SM also acknowledges partial financial contributions from the STFC Consolidated Grant ST/L000296/1. Furthermore, the work of LDR has been supported by the STFC/COFUND Rutherford International Fellowship scheme. 
SK was partially supported by the STDF project 13858 and the European Union's Horizon 2020 research and innovation programme under the Marie Sklodowska-Curie grant agreement No. 690575.
All authors finally acknowledge support from the grant H2020-MSCA-RISE-2014 No. 645722 (NonMinimalHiggs).

\providecommand{\href}[2]{#2}\begingroup\raggedright\endgroup

\end{document}